# FeederGAN: Synthetic Feeder Generation via Deep Graph Adversarial Nets

Ming Liang, *Student Member, IEEE*, Yao Meng, *Student Member, IEEE*, Jiyu Wang, *Student Member, IEEE*, David Lubkeman, *Fellow*, *IEEE*, and Ning Lu, *Senior Member*, *IEEE*

*Abstract*— This paper presents a novel, automated, generative adversarial networks (GAN) based synthetic feeder generation mechanism, abbreviated as FeederGAN. FeederGAN digests real feeder models represented by directed graphs via a deep learning framework powered by GAN and graph convolutional networks (GCN). Information of a distribution feeder circuit is extracted from its model input files so that the device connectivity is mapped onto the adjacency matrix and the device characteristics, such as circuit types (i.e., 3-phase, 2-phase, and 1-phase) and component attributes (e.g., length and current ratings), are mapped onto the attribute matrix. Then, Wasserstein distance is used to optimize the GAN and GCN is used to discriminate the generated graphs from the actual ones. A greedy method based on graph theory is developed to reconstruct the feeder using the generated adjacency and attribute matrices. Our results show that the GAN generated feeders resemble the actual feeder in both topology and attributes verified by visual inspection and by empirical statistics obtained from actual distribution feeders.

## I. INTRODUCTION

Test systems are widely used by researchers and engineers to test conceptual designs, optimize parameter settings, and validate performance [1–6]. A test system can be either a real feeder model or a generalized, derived, or extended versions of an actual feeder. For example, a number of new test systems can be obtained by varying the network topology, adjusting the placement and parameters of apparatus (e.g., adding new devices, removing old ones, retrofitting, etc.), and changing user patterns in an actual utility feeder model. A realistic test system needs to faithfully reproduce important distribution system behaviors in different operation conditions so that researchers and engineers can benchmark their algorithms and conduct studies on a wide range of operation conditions with less development time, lower testing cost, and zero risk to interrupt actual system operation.

In recent years, the needs for such realistic distribution feeder test systems are increasing drastically because of the integration of distributed energy resources (DERs) [7-8], energy storage systems [9-10] and electric vehicles (EVs) [11-12]. To model the uncertainty and variability in DER operation as well as the variations in its operation environment (e.g., changes in topology, variation in model parameters, user patterns, etc.), an ensemble of realistic distribution test feeders are often needed.

However, developing high-fidelity distribution feeder models requires access to utility network models and customer data, which is a major barrier for the research community to have unrestrictive, unlimited number of customizable, realistic test systems for research and development purpose. Meanwhile, domain experts, such as operation and planning engineers, have conventions to follow and specific design criterions to meet when designing actual engineering systems. Those inherent characteristics are very hard to be captured by researchers even when they can access actual system models.

Since 1991, 10 IEEE test feeders were developed to meet researcher's needs, approximately. In 2009, led by a group of researchers at Pacific Northwest National lab, 24 taxonomy feeders were developed from 575 actual feeders from 17 utilities across the US [3], [13]. Schneider et al. summarize the original intent for different test feeders and help researchers make decision on which of the test feeders are most appropriate for their work [14]. At Texas A&M, researchers are focusing on developing synthetic electric grid models that are designed to be statistically and functionally similar with actual electric grids while containing no confidential critical energy infrastructure information [4].

However, so far, there has been very little attempt made towards the manual and static test system design principles, making creating an ensemble of test systems from actual feeder models a daunting task. In recent years, researchers develop a few methods that use street maps to align and generate feeder layouts [15–17]. For example, in [16], Saha et al. use an open source street map tool to create individual synthetic distribution feeders and groups of feeders in an area with the same ZIP code using a small number of input data, while in [17], Domingo et al. use a reference model to generate large-scale radial feeders using street maps as inputs. In [18], [19], Birchfield et al. summarize topological and statistical criterion to validate the realism of synthetic feeders. In [20], Shweitzer et al. conduct a two-stage study to generate synthetic feeders where they treat the feeder as a graph with nodes and edges. In the first stage, a thorough analysis is conducted to summarize the topological and physical characteristics where divergence is used to measure the statistical similarity of real and synthetic feeders. In the second stage, summarized statistics of the feeders are used to generate synthetic feeders having similar characteristics with those Netherland feeders. The shortcoming of these approaches is that they are semi-automatic and lack of the

This study is funded by North Carolina State University. Ming Liang, Yao Meng, Jiyu Wang, David Lubkeman and Ning Lu are with the Electrical & Computer Engineering Department, Future Renewable Energy Delivery and Management (FREEDM) Systems Center, North Carolina State University, Raleigh, NC 27606 USA. (e-mails: mliang3@ncsu.edu, ymeng3@ncsu.edu, jwang49@ncsu.edu, dllubkem@ncsu.edu, nlu2@ncsu.edu).

intelligence to learn feeder typology or devices attributes from ingesting actual feeder models, making them unsuitable for generating a large amount of realistic, up-to-date test feeders.

In lieu of this, our aim is to develop an end-to-end, machine-learning-based approach for automated, customizable test feeder generation using actual feeder models as inputs. In [21], Kipf et al. propose the graph convolutional networks (GCN) for classifying graph data, such as social network and chemical molecules. Based on GCN, many follow-up studies choose to use a generative adversarial networks (GAN) with GCN to generate graph data. For instance, in [22], Fan and Huang use labeled graph in GAN to generate citation graph and protein graph. A recurrent network based model is used in [23] by You et al. to generate synthetic graphs.

Motivated by those approaches, we develop a GAN-based synthetic feeder generation mechanism, abbreviated as FeederGAN. FeederGAN digests real feeder models using a deep learning framework powered by GAN and graph convolutional networks (GCN). From the information provided in the input files of a power system feeder model, we first map the device connectivity to the adjacency matrix, and then map the device characteristics such as circuit types (i.e., 3-phase, 2-phase, and 1-phase) and component attributes (e.g., length and current ratings) to the attribute matrix, making GCN training possible. Wasserstein distance is used to optimize the GAN for discriminating the generated graph from the actual. We also introduce a method for reconstructing feeder topology from the obtained adjacency and attribute matrix based on graph theory.

Our contributions are three-fold. *Frist*, the topology is learnt from ingesting actual feeder models instead of street maps, making it possible to extract inherent distribution system design features. *Second*, to our best knowledge, we are the first to use a GAN-based end-to-end and plug-to-play model for generating synthetic feeders. *Third*, FeederGAN is scalable and customizable so its user can generate either a whole feeder or substructures of a feeder at user selected scale and size.

## II. GRAPHICAL REPRESENTATION OF A FEEDER

In this section, we introduce two main considerations when converting a power system feeder model to a machine-learning friendly directed graph model and then discuss the problem formulations of the directed graph model.

The *first* consideration is to omit the geographical information and use the length of the feeder line to represent distance. This is because, in power system network models, the electrical distance used for calculating line impedance is of interest instead of the geographical distance. For visualization purpose, a greedy method (introduced in Section VI.D) is developed to calculate pseudo coordinates for each bus. The substation (i.e., the head of a feeder) is placed at the origin (0, 0). Then, coordinates of subsequent buses are calculated one by one based on electrical distance and connectivity so that the generated synthetic feeder stretches forward as straight as possible, making it easy to traverse feeder topology and identify the relationships between feeder components. As shown in Fig. 1, this approach removes the need for using sensitive customer geographic information, such as street maps and geographic coordinates. From the machine learning standpoint, masking geographical information forces the algorithm to use only the relative distance between line segments, preventing the algorithm from learning from non-generic, geographic information.

The *second* consideration is to replace the traditional bus-as-node and device-as-edge feeder topology representation with a device-as-node representation. Using the device-as-node approach, each device (e.g., a line segment instead of a bus) is defined as a node so the edge will only serve as an indicator to show the direction pointing from the feeder head to a load node. By doing so, the device attributes, such as length, current ratings, phase, can be defined as node attributes.

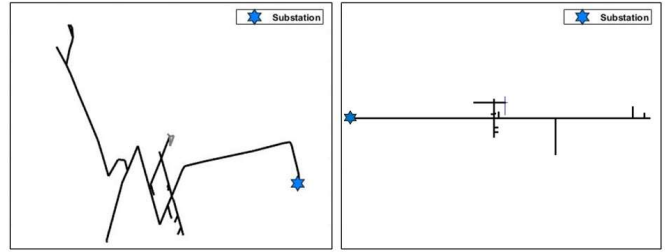

Fig.1 Topology representation: feeder using geographical coordinates (left) and using electrical distance (right)

The device-as-node representation allows us to represent each device by a node $v_i \in \mathcal{V}$ and a node attribute vector $\mathbf{x}_i$. The feeder can be represented as a directed graph $\mathcal{G}$ with a set of nodes $\mathcal{V}$ and edges $\mathcal{E}$ so that each edge $(v_i, v_j) \in \mathcal{E}$ connects two nodes and shows the graph direction. Then, the feeder topology can be represented by a directed graph by an adjacency matrix $\mathbf{A} \in \mathbb{R}^{m \times m}$ and an attribute matrix $\mathbf{X} = [\mathbf{x_1}, ..., \mathbf{x_m}]^T \in \mathbb{R}^{m \times d}$, where $m$ is the number of nodes, $\mathbf{A}_{i,j} \in \{0, 1\}$ and $d$ is the dimension of the attributes. After converting the feeder topology to a directed graph, the learning algorithm only needs to learn from $\mathbf{A}$ and $\mathbf{X}$, significantly reducing the problem complexity, as shown in Fig. 2.

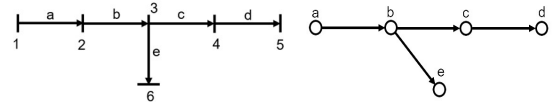

Fig. 2 Distribution feeder topology representation: Bus-as-node and device-as-edge (left) and device-as-node (right)

**Assumption 1:** Because traditional distribution feeders are designed as radial networks, we assume that the feeder graph is a directed radial graph and there is no loop in the graph.

**Property 1:** If Assumption 1 holds, the in-degree is 1 for all nodes except the feeder head node, whose in-degree is 0. This property suggests that there should be one and only one non-zero element in each column of the adjacency matrix $\mathbf{A}$, except for the column representing the feeder head.

## III. PROBLEM FORMULATION

This section introduces the deep learning concepts used in the paper, the preparation of training data, and the problem formulation of GAN and GCN for generating synthetic feeder topology and attributes.

### A. Flowchart of the GAN-based Framework

The flowchart of the FeederGAN framework is introduced in Fig. 3. First, in the "format real feeder as graphs step", the topology and attribute information are extracted from the input files of actual distribution feeder models to obtain the actual adjacency matrix **A** and attributes matrix **X**. Then, we will train the Generator and Discriminator of the GAN, in which Discriminator will compare the Generator generated $\widehat{\mathbf{A}}$ and $\widehat{\mathbf{X}}$ with the actual **A** and **X** until they can no longer be distinguished (i.e. converged). Then, a greedy method based on graph theory is developed to reconstruct the feeder from the generated $\widehat{\mathbf{A}}$ and $\widehat{\mathbf{X}}$. After that, we will check for the quality of the GAN-generated feeders. Details of each step will be described in subsequent sections.

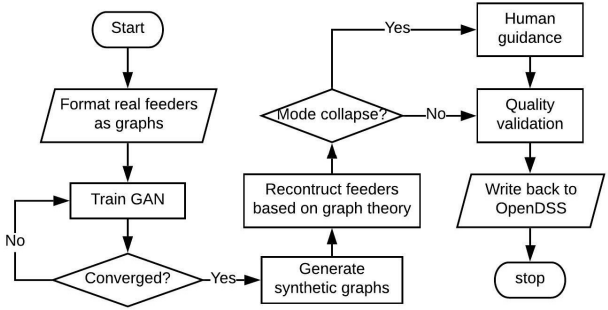

Fig. 3. Flowchart of FeederGAN

### B. An Introduction of the Deep Learning Framework

FeederGAN digests real feeder models using a deep learning framework powered by GAN and GCN. Deep learning refers to deep neural networks built on several layers of neutrons. If the layer is linearly and fully connected, it is called a fully connected (FC) layer. A model built on several FC layers is called multilayer perceptron (MLP). To model the non-linearity in the dataset, activation functions are used. In this paper, two major activation functions are rectified linear units (ReLU) and softmax. ReLU [24] is formulated as $f(x) = \eta x$, where $\eta = 1$ when $x > 0$; $\eta = 0$ when $x \leq 0$. If the negative part $\eta$ is a small but non-zero value, the method is called leaky ReLU [24]. Softmax [25] is formulated as $f(x_i) = e^{x_i} / \sum_j e^{x_j}$. Softmax is often used in classification or approximating discrete one-hot values, where a discrete variable is encoded as a vector consisting of binary values (i.e., 0/1). When using Softmax, the output of each layer is normalized as a probability and the sum of all outputs is one.

### 1) Generative Adversarial Networks (GAN)

As shown in Fig. 4, a GAN [26] consists of two main components: Generator and Discriminator. Discriminator distinguishes whether the input data is real or fake (i.e., generated) while Generator generates fake data sets resembling the real data to exploit weakness in Discriminator. Thus, GAN learns to generate high-quality fake data sets by letting Generator and Discriminator play a minimax game, which ends when Discriminator can no longer distinguish the generated and real data. This process can be formulated as

$$\min_G \max_D V = \mathbb{E}_x[\log(D(x))] + \mathbb{E}_z[\log(1 - D(G(z)))] \quad (1)$$

where $D(x)$ is Discriminator outputs, $G(z)$ represents Generator output, $x$ represents the real data, and $z$ represents a random noise. For Discriminator, the first part of (1) is to predict the real as real, and the second part is to predict the fake as fake. $D(x)$ outputs are interpreted as probability with 1 for real and 0 for fake. To make the Discriminator output a probability, a sigmoid function [26] is added to the

Discriminator output layer to normalize $D(x)$ into [0, 1]. For Generator, the first part of (1) is not used because the Generator goal is to make Discriminator "think" the fake is real.

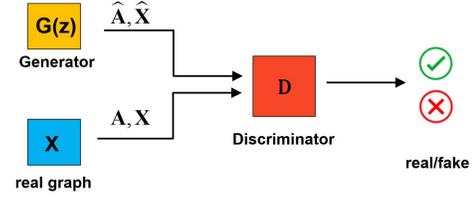

Fig. 4. Framework of a typical GAN

### 2) Graph Convolutional Networks (GCN)

GCN [21] uses the adjacency matrix **A** to represent the connectivity between nodes and the attribute matrix **X** to store nodal device information. The GCN representation of a directed graph is

$$\bar{\mathbf{A}} = \mathbf{D}^{-1}(\mathbf{A} + \mathbf{I}) \quad (2)$$

$$\text{GCN}(\bar{\mathbf{A}}, \mathbf{X}) = \sigma(\bar{\mathbf{A}}\mathbf{X}\mathbf{W}) \quad (3)$$

where **I** is an identity matrix, **D** is a diagonal degree matrix with $\mathbf{D}_{ii} = 1 + \sum_j \mathbf{A}_{ij}$, and **W** is the network parameter matrix to be learnt. $\sigma(\cdot)$ is an activation function (e.g., ReLU). Note that by using (2), a row-wise normalization is executed to **A**. To simplify future derivations, we let **A** be the normalized adjacency matrix in the rest of the paper.

### C. Feature Engineering

Feeder topology information, i.e., the connectivity between feeder nodes, are coded into **A**, a $m \times m$ matrix. Let $\mathbf{A}_{i,j} = 1$ if there is a connection from node $i$ to $j$ (unilateral) and let $\mathbf{A}_{i,j} = 0$ otherwise. After the whole network is traversed, use (2) to normalize **A**.

As shown in Table I, there are two types of features in **X**: organic and topological. Organic features include device length, current rating, and phase, the information of which can be directly extracted from device definition files. Topological features include the distance between the device and the feeder head, pseudo loads the device carries, and the device level in the overall feeder topology hierarchy. Topological features are obtained by traversing the feeder topology using either the depth-first search (DFS) or the breadth-first search (BFS).

Table I A summary of the attributes
$\mathcal{O}$: organic, $\mathcal{T}$: topological, $\mathcal{N}$: numerical, $\mathcal{C}$: categorical

| Name | Definition | Type | Source |
|------|------------|------|--------|
| Length | The length of a device. | $\mathcal{N}$ | $\mathcal{O}$ |
| Norm Amps | Normal condition conductor amps, an indicator for the conductor materials. | $\mathcal{N}$ | $\mathcal{O}$ |
| Distance | Distance from feeder head to the device. | $\mathcal{N}$ | $\mathcal{T}$ |
| Pseudo Load | The sum of the capacity of all downstream customer side transformers. | $\mathcal{N}$ | $\mathcal{T}$ |
| Level | Start as Level 0 at the feeder head. When encountered a bifurcation leading to several children branches, level+1 if "norm amps" or "phase" of the child is different from that of the parent. | $\mathcal{C}$ | $\mathcal{T}$ |
| Phase | 1 of the 7 options: *a, b, c, ab, ac, bc, abc* | $\mathcal{C}$ | $\mathcal{O}$ |

In Table I, the first four features have continuous, numerical values and are normalized by their extreme values to uniformly map them into [-1, 1]; the last two features are categorical features that bear discrete values and are encoded by one-hot values. Thus, numerical attributes are represented by $\mathbf{X_{num}} \in$ $\mathbb{R}^{m \times 4}$ and categorical attributes by $\mathbf{X}_{cat} = [\mathbf{X}_{cat}^1, \mathbf{X}_{cat}^2] \in \mathbb{R}^{m \times (d_1 + d_2)}$, where $m$ is the number of nodes, and $d_1, d_2$ are the dimension of each categorical feature, i.e., the number of categories for each feature. FeederGAN Algorithm and Training Strategy

In our experiments, the minimax GAN problem formulated in (1) suffers convergence issues caused by gradient vanishing, where the loss of Discriminator decreases quickly when reaching to an optimal solution, causing backward gradients of the Generator to drop to 0. As a result, the Generator can no longer be trained properly. To solve this problem, we apply the Wasserstein GAN (WGAN) introduced by Arjovsky in [27]. WGAN shares the same model structure with GAN, but by removing the sigmoid output layer of the GAN, WGAN directly outputs a logit, called the critic score, instead of a probability. Thus, the gap of the critic score between a real graph and a generated graph, called the Wasserstein distance, can be used to train the Generator and Discriminator.

The WGAN formulation is

$$\min_G \max_D V = \mathbb{E}_x[D(x)] - \mathbb{E}_z[D(G(z))] \quad (4)$$

where $x$ is the real graph, i.e. $\{\mathbf{A}, \mathbf{X}_{num}, \mathbf{X}_{cat}\}$, $z$ is a random noise matrix, and $G(z)$ is the generated graph $\{\hat{\mathbf{A}}, \hat{\mathbf{X}}_{num}, \hat{\mathbf{X}}_{cat}\}$, $\mathbb{E}_x[D(x)]$ and $\mathbb{E}_z[D(G(z))]$ is the critic score the Discriminator gives to the real graph and the generated graph, respectively. Now, the Discriminator's objective is to widen the Wasserstein distance while the Generator's objective is to narrow it.

When there is no constraint on the Discriminator, the value of the Discriminator weights will gradually grow larger and larger and thus fail to train Generator properly. Hence, the Discriminator will be subject to a constraint called the $k$-Lipschitz condition such that

$$|D(x_1) - D(x_2)| < k \times |x_1 - x_2| \quad (5)$$

To implement the Lipschitz condition, we conduct weights clipping so that the weights of Discriminator are clamped within $(-c, c)$, where $c$ is the clipping limit. The FeederGAN algorithm is summarized in Algorithm 1.

**Algorithm 1.** FeederGAN. All experiments in the paper used the default training parameters $\alpha=0.0001$, $c=0.1$, $n=20$, $n_0=500$, $n_1=1000$, $n_2=5$.

**Require:** $\alpha$, the learning rate; $c$, the weight clipping limit; $n_1, n_2$, the number of iterations for Discriminator at different stages; $n$, $n_0$, the numbers separate different training stages; $\omega_0$, initial Discriminator parameters; $\theta_0$, initial Generator parameters.

1:   $i = 0$
2:   **while** $\theta$ not converge **do**
3:      $iter \leftarrow n_1$, **if** $i < n$ **or** $i \% n_0 = 0$
4:      $iter \leftarrow n_2$ **else**
5:      **for** $t=1, 2, \ldots, iter$ **do**
6:         Draw a sample $\{\mathbf{A}, \mathbf{X}_{num}, \mathbf{X}_{cat}\}$ from the real dataset
7:         Generate a random matrix $\mathbf{z}$ from the prior distribution
8:         $\{\hat{\mathbf{A}}, \hat{\mathbf{X}}_{num}, \hat{\mathbf{X}}_{cat}\}_\theta \leftarrow G_\theta(\mathbf{z})$
9:         $g_\omega \leftarrow \nabla_\omega[D_\omega(\{\mathbf{A}, \mathbf{X}_{num}, \mathbf{X}_{cat}\}) - D_\omega(\{\hat{\mathbf{A}}, \hat{\mathbf{X}}_{num}, \hat{\mathbf{X}}_{cat}\}_\theta)]$   (6)
10:        $\omega \leftarrow \omega + \alpha \cdot \text{RMSProp}(\omega, g_\omega)$
11:        $\omega \leftarrow clipping(\omega| -c, c)$
12:      **end for**
13:      Generate a random matrix $\mathbf{z}$ from prior distribution
14:      $\{\hat{\mathbf{A}}, \hat{\mathbf{X}}_{num}, \hat{\mathbf{X}}_{cat}\}_\theta \leftarrow G_\theta(\mathbf{z})$
15:      $g_\theta \leftarrow -\nabla_\theta[D_\omega(\{\hat{\mathbf{A}}, \hat{\mathbf{X}}_{num}, \hat{\mathbf{X}}_{cat}\}_\theta)]$   (7)
16:      $\theta \leftarrow \theta - \alpha \cdot \text{RMSProp}(\theta, g_\theta)$
17:      $i \leftarrow i + 1$
18:   **end while**

Lines 3-4 illustrate the strategy used to train Discriminator as strong as possible. As shown in Fig. 5, in the first few iterations, the Generator will be trained once every $n_1$ Discriminator training cycles. After the Generator has been trained $n$ times, the frequency will be reduced to train Generator once every $n_2$ Discriminator training cycles. Then, at the $n_0, 2n_0, 3n_0, \ldots$ Generator training cycles, we will enforce an aggressive training snapshot to train the Discriminator $n_1$ times, before training the Generator. By doing so, we can train the Discriminator as strong as possible to widen the Wasserstein distance, making it easy to distinguish the real graph from the generated ones. If the Generator can be trained to narrow the Wasserstein distance, the generated graphs eventually will be realistic enough to cheat the Discriminator.

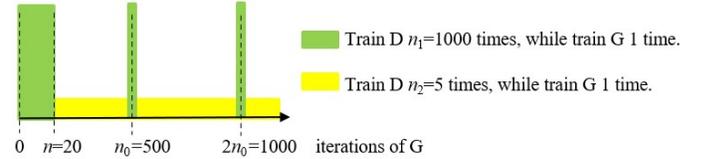

Fig. 5. FeederGAN training schedule

Lines 5-12 illustrate the Discriminator training process. At each training session, a real graph combined with a generated graph are used to obtain the Wasserstein distance. Then, we fix the parameters of the Generator and trace backward to calculate the gradients using (6). After that, we use the RMSProp algorithm [28] to adaptively update the weights of the Discriminator. In the end, we conduct a weight clipping so that for weights smaller than $-c$, set them at $-c$; for weights larger than $c$, set them at $c$. Similarly, lines 13-16 describe how to calculate the Generator gradients and update its weights.

## IV. FEEDERGAN IMPLEMENTATION

This section introduces the architecture of the Generator and the Discriminator as well as hyperparameters.

### A. Generator Architecture

As shown in Fig. 6, a multilayer perceptron (MLP) that has several FC layers is used to generate $\hat{\mathbf{A}}$ and $\hat{\mathbf{X}}$. The random noise matrix $\mathbf{Z}$ is generated using a prior Gaussian distribution and the latent matrix $\mathbf{L}$ is generated using FC1. FC2 is used to generate the numerical attributes $\hat{\mathbf{X}}_{num}$ and a hyper tangent activation function is used to scale $\hat{\mathbf{X}}_{num}$ within [-1, 1], the same value range as the real attributes. The process is described as

$$\mathbf{L} = \text{FC1}(\mathbf{Z}), \hat{\mathbf{X}}_{num} = \text{FC2}(\mathbf{L}), \hat{\mathbf{X}}_{num} = \tanh(\hat{\mathbf{X}}_{num}). \quad (8)$$

To generate the categorical features, we first use FC3 to generate $\mathbf{V}$, the latent matrix for all categorical features. Then, a separate FC4-$i$ layer is used to generate the $i^{th}$ categorical feature. After that, a row-wise softmax activation function is used to approximate the one-hot encoded categorical feature before outputting it to $\hat{\mathbf{X}}_{cat}^i$. The use of *softmax* makes the network differentiable. Note that if a hard encoding method that is not non-differentiable (e.g., *argmax*) is used, the gradient backward pass will be broken. This process for generating the categorical features is described as

$$\mathbf{V} = \text{FC3}(\mathbf{L}), \hat{\mathbf{X}}_{cat}^i = \text{FC4} - i(\mathbf{V}), \hat{\mathbf{X}}_{cat}^i = \text{softmax}(\hat{\mathbf{X}}_{cat}^i) \quad (9)$$

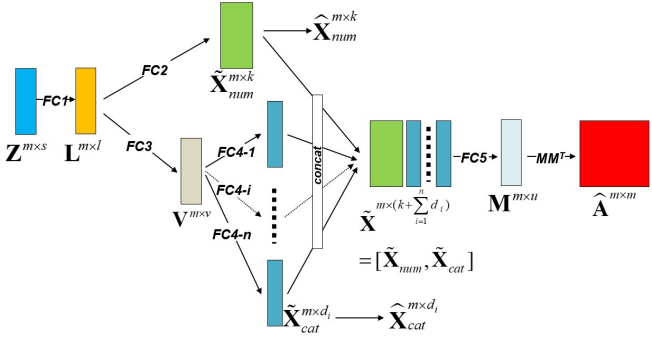

Fig. 6. Generator architecture

The numerical and categorical features (before activation) are then concatenated to form the latent matrix $\widetilde{\mathbf{X}}$. Since the two features have different value scales, they will first be normalized then go through FC5 to generate $\mathbf{M}$. $\mathbf{M}$ is then multiplied by its transpose to form an $m \times m$ matrix, which will then go through a column wise softmax activation function to obtain $\widehat{\mathbf{A}}$. Using column-wise softmax, we can map all $\widehat{\mathbf{A}}$ entries to $[0, 1]$ while enforcing Property 1 (i.e., only 1 non-zero entry in each column), making those potentially non-zero entries stand out. This process is formulated as

$$\widetilde{\mathbf{X}} = \text{norm}\left(\left[\; \widetilde{\mathbf{X}}_{num}, \widetilde{\mathbf{X}}_{cat}\right]\right), \mathbf{M} = \text{FC5}(\widetilde{\mathbf{X}}), \widehat{\mathbf{A}} = \text{softmax}(\mathbf{M} \times \mathbf{M}^T)\quad(10)$$

### B. Discriminator Architecture

The objective of the Discriminator is to distinguish the real and generated graphs. The discriminative process for both the real and generated graphs is the same, as shown in Fig. 7. Hence, only the generated set of data $\widehat{\mathbf{A}}$, $\widehat{\mathbf{X}}_{num}$, $\widehat{\mathbf{X}}_{cat}$ are used to illustrate the discriminative process.

Categorical features are either one-hot encoded or softmax approximated, so an embedding layer emb-$i$ is used to project those discrete values into a lower dimension matrix, $\mathbf{P}_i$. Thus, the embedding process not only conducts dimension reduction, but also maps the discrete, sparse values into continuous, dense values, making it possible for the categorical features to be processed together with the numerical features.

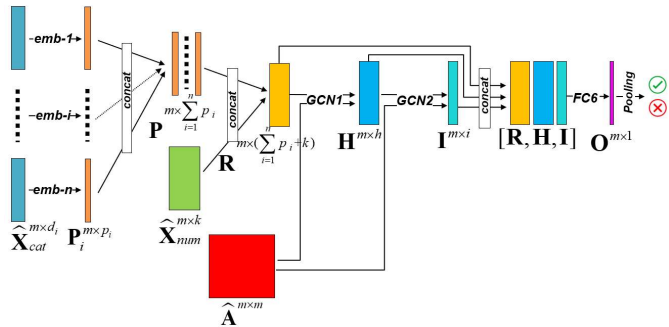

Fig. 7. Discriminator architecture

Symmetrically as in the generative process, all the embedding results are concatenated as a latent representation, $\mathbf{P}$, for categorical features. Then, $\mathbf{P}$ is concatenated with the numerical features and go through a layer normalization to get the latent matrix $\mathbf{R}$ for all the attributes. The attributes processing is formulated as

$$\mathbf{P}_i = \text{emb} - i(\widehat{\mathbf{X}}_{cat}^i), \mathbf{R} = \text{norm}([\mathbf{P}, \widehat{\mathbf{X}}_{num}])\quad(11)$$

Next, we use the adjacency matrix to filter the attributes via two GCN layers. Since GCN is prone to gradient vanishing problem, a residual structure is used to aggregate the latent matrices $\mathbf{R}, \mathbf{H}, \mathbf{I}$ together. Hence, the gradient backward pass becomes a three-way pass instead of a single one to boost the gradients back and benefit the training of the Generator.

In the second to the last layer, FC6 processes the concatenated residual matrices to obtain a vector $\mathbf{O}$ with $m$ entries that reveals the latent information for each node. Finally, a global pooling method, e.g. max-pooling, is used to output a single value serving as the critic score. The discriminative process is formulated as

$$\mathbf{H} = \text{GCN1}(\widehat{\mathbf{A}}, \mathbf{R}), \mathbf{I} = \text{GCN2}(\widehat{\mathbf{A}}, \mathbf{H})\quad(12)$$
$$\mathbf{O} = \text{FC6}([\mathbf{R}, \mathbf{H}, \mathbf{I}]), \text{score} = \text{pooling}(\mathbf{O})\quad(13)$$

Note that besides softmax and hyper tangent, we use ReLU after certain FC layers and leaky ReLU after GCN to introduce non-linearity.

### C. Hyperparameters

The major hyperparameters are listed in the first row of Algorithm 1. To guarantee the convergence of Wasserstein distance, experiments need to be conducted to tune the learning rate. In our paper, we choose a learning rate $\alpha$=0.0001. The clipping limit $c$ is also determined through experiments and is set at 0.1 to avoid gradient vanishing and exploding.

The dimension of the random noise matrix $\mathbf{Z}$ is $m \times s$, where $m$ is the number of nodes that is adaptive to the real graphs; $s$ is the parameter to be tuned, and we set as 20 in this paper. The size of the hidden layers is subject to the size of the real graphs where typical values of 128 and 64 are used.

## V. FEEDER RECONSTRUCTION

Once the training of Generator finished, we can use it to generate $\widehat{\mathbf{A}}$ and the attribute matrices $\widehat{\mathbf{X}}_{num}$, $\widehat{\mathbf{X}}_{cat}$, which are either softmax-approximated to be within $[0, 1]$ or hyper tangent approximated to be within $[-1, 1]$. Then, we use the reconstruction methods to map them back to the actual values.

### A. Reconstruction of the Adjacency Matrix

The entries in the generated adjacency matrix $\widehat{\mathbf{A}}$ have continuous values in $[0, 1]$. Algorithm 2 is used to reconstruct the binary directed adjacency matrix for the feeder from $\widehat{\mathbf{A}}$.

**Algorithm 2.** Adjacency Matrix Reconstruction.

**Require:** $\widehat{\mathbf{A}}$, the generated soft adjacency matrix; *row,* the number of rows in the adjacency matrix; *col,* the number of columns in the adjacency matrix.

1:   **for** $i=1, 2, \ldots, row$ **do**
2:     **for** $j=i, i+1, \ldots, col$ **do**
3:       $\widehat{\mathbf{A}}_{j,i} \leftarrow 0$, if $\widehat{\mathbf{A}}_{i,j} \geq \widehat{\mathbf{A}}_{j,i}$
4:       $\widehat{\mathbf{A}}_{i,j} \leftarrow 0$, if $\widehat{\mathbf{A}}_{i,j} < \widehat{\mathbf{A}}_{j,i}$
5:     **end for**
6:   **end for**
7:   $\widetilde{\mathbf{A}} \leftarrow \widehat{\mathbf{D}}^{-1}\widehat{\mathbf{A}}$
8:   **for** $j=1, 2, \ldots, col$ **do**
9:     $k \leftarrow \text{argmax}(\widetilde{\mathbf{A}}_{:,j})$, $\widetilde{\mathbf{A}}_{k,j} \leftarrow 1$
10:     **for** $i=1, \ldots, row$ and $i \neq k$ **do**
11:       $\widetilde{\mathbf{A}}_{i,j} \leftarrow 0$
12:     **end for**
13:   **end for**
14:   **return** $\widetilde{\mathbf{A}}$

In lines 1-6, to make a directed graph out of $\widehat{\mathbf{A}}$, for any symmetrical pair $\widehat{\mathbf{A}}_{ij}$ and $\widehat{\mathbf{A}}_{ji}$, there should be one non-zero entry at most. Hence, we set the smaller one as 0 and keep the larger one as it is. In line 7, we conduct a row-wise normalization, which is also used in the feature engineering process as an import procedure to guarantee sparsity. In lines 8-13, we conduct Property 1, i.e. in each column we set the largest entry as 1 and the rest as 0. After that, $\tilde{A}$ becomes a binary, sparse adjacency matrix describing a directed graph.

### B. Topology Reconstruction

In graph theory, node ordering can be problematic, because for a specific adjacency matrix, the different ordering of nodes can represent different topologies. For the Discriminator, because GCN is ordering invariant, node ordering is not an issue. However, for Generator, MLP is ordering variant, so node ordering needs to be addressed. Note that using permutation to resolve the node ordering is impossible as a graph with $m$ nodes has $m$-factorial ($m!$) possibilities.

To reduce the complexity, one can permutate only the position of the feeder head based on the reconstructed $\tilde{A}$, which has only 1 non-zero entry in each column. The algorithm is illustrated in Algorithm 3. The node-to-edge transformation (in line 5) is an inverse procedure of Fig. 2. Therefore, for each generated adjacency matrix, theoretically, $m$ topologies can be obtained at the most. For more detailed explanations, please refer to Example 1 in Appendix.

---

**Algorithm 3.** Feeder Head Permutation.

**Require:** $\tilde{A}$, the reconstructed binary directed matrix; $m$, the node number.

1: **for** $i = 1, 2…, m$ **do**
2:      **if** row $i$ with non-zero entries **do**
3:          Set the corresponding column $i$ as all 0s;
4:          Let node $i$ be the feeder head and traverse the adjacency matrix to build the graph;
5:          Conduct a node-to-edge transformation to build the feeder topology.
6:      **end if**
7: **end for**

---

### C. Attributes Reconstruction

After reconstructing the feeder topology, we need to assign attributes to each device so that a distribution feeder model in the targeted format (e.g., OpenDSS, CYME) can be produced. Among all attributes, topological features are just used to facilitate the discriminative inference, which are not necessary to be reconstructed. Hence, we only need to reconstruct the organic features.

For "length", we map the generated value from [-1, 1] to [0, 800] uniformly, where 800 m is the nominal maximum value of a feeder line segment. Note that one can pick a suitable maximum value based on his modeling need. Although the "norm amps" is used as a continuous value, it's used to represent the conductor material information, e.g. line code, thus the category of the values is limited to about 10. Hence, based on the generated value, we use a lookup table for line codes to assign the nearest "norm amps" value. Among all attributes, "phase" is the most important one because it affects how realistic a generated feeder is. As *softmax* is used to approximate the one-hot encoding, an *argmax* function is used to map the phase type to the 7 choices.

### D. Visualize the Generated Feeder

As the geographical coordinates have been omitted in the learning process, we need to generate pseudo coordinates for each bus to plot out the feeder topology for inspection. A greedy method (as described in Algorithm 4) is developed to generate the pseudo coordinates with two objectives: making the feeder lines as straight as possible while reducing the chance of overlapping and crossing. For more details, please refer to Example 2 in Appendix.

---

**Algorithm 4.** Feeder Topology Visualization

1: Set the feeder head as the origin (0,0) and traverse the rest of the feeder;
2: Set the direction as straight right and make the direction angle as $\theta = 0$;
3: If no bifurcation, keep the current direction, calculate the coordinate of the next bus using the line length;
4: When there is a bifurcation, rank the priority of children branches based on the number of its downward devices. The large the number, the higher the ranking;
5: The 1st child branch keeps the same direction as parent's, the 2nd and 3rd children's angle plus a delta angle $\delta$ ($\pi/2$) and (-$\pi/2$) respectively. If there are more children in the branch, use more angles, e.g. ($\pi/4$), ($3\pi/4$), ($-\pi/4$), ($-3\pi/4$), ($\pi/8$), ...;
7: Update the children branch's direction angle, calculate the corresponding bus coordinates with the angle, the length of the line, and the parents' coordinates.

---

## VI. RESULTS AND DISCUSSION

### A. Dataset and Augmentation

We use real feeder models with 1500 to 2000 overhead lines and cables as inputs in this study. First, all tie switches are set as "open" to satisfy Assumption 1 (i.e., no loops). Second, we sample sub-feeder structures from full feeder networks to augment the training database using Algorithm 5. Then, the full feeder network and the sampled sub-feeders are used as the training dataset, in which we have 664 graphs in total.

---

**Algorithm 5.** Subgraph Sampling.

**Require:** A pool of real feeders served as real graphs

1: Select a start node from level 0 or level 1 nodes;
2: Extract all its downstream nodes (all the way to loads) to form a subgraph;
3: If number of nodes is less than 100, resample;
4: If the number of nodes is more than 50% of the original graph, resample;
5: Repeat for desired times.

---

### B. Training Loss and Generated Feeders

In Fig. 8, the training losses of the Discriminator when feeding with real and generated (fake) graphs, and the Wasserstein distance are shown on the left; the zoom-in plots for each training stage are shown on the right. The x-axis is the number of iterations for Generator. Note that one Generator iteration corresponds to multiple Discriminator iterations.

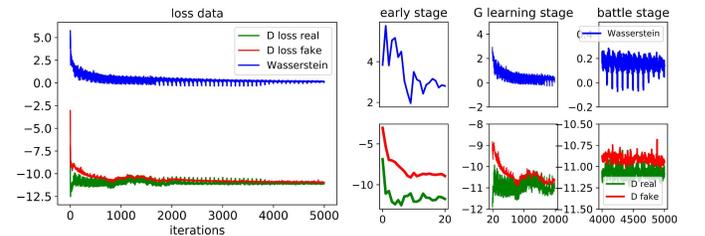

Fig. 8. Loss and Wasserstein distance

In the first few iterations, we train Discriminator to be very strong so it can easily widen the Wasserstein distance. When the training proceeds, although the Discriminator is trained more times than the Generator, the Generator learning process will gradually narrow the Wasserstein distance until the distance converges to a very small but non-zero oscillatory equilibrium, as shown in the zoom-in plots. Meanwhile, the critic score for the real and the fake vibrates around an equilibrium point. It shows that the Generator generates graphs of good quality that the Discriminator can no longer tag them.

Figure 9 shows examples of the feeder topologies generated by the FeederGAN, from which, an expert can examine the feeder layout and exclude the ones that do not meet their requirements. Note that we only generate line segments, i.e., overhead lines or cables in current stage. Reclosers or regulators can be added later based on different modeling needs by the user.

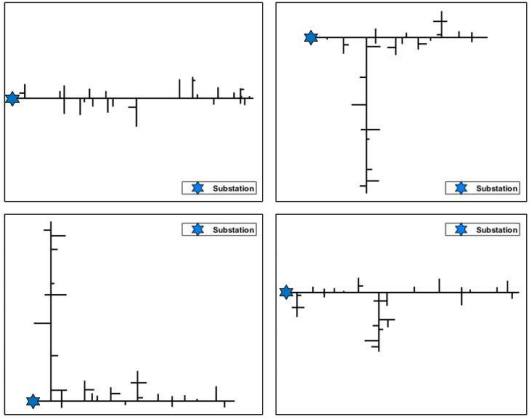

Fig. 9. Generated feeders

Once the generation process is finished, the FeederGAN tool automatically converts feeder topology and attributes back into OpenDSS input file formats for running power flow studies. Figure 10 shows an example of the nodal voltage distribution along the FeederGAN-generated feeder and Fig. 11 shows the voltage profile of each phase with respect to the distance to the substation.

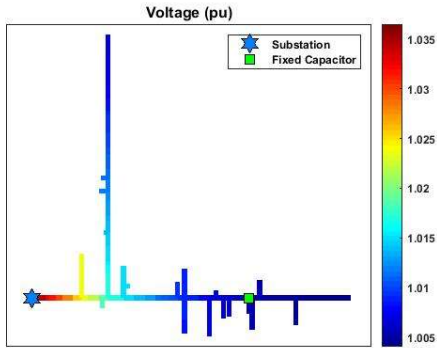

Fig. 10. Nodal voltage along the generated feeder

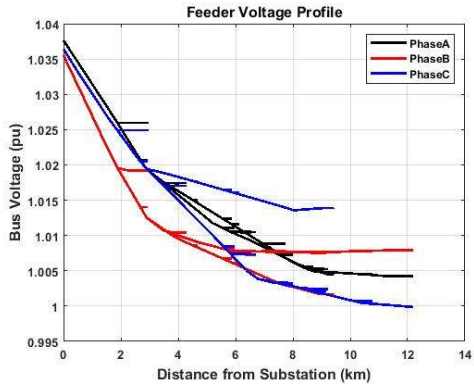

Fig. 11. Voltage profiles of nodes on phase a, b, and c (Nodes arranged ascendingly according to their distance to the substation)

## C. Performance Metrics

We define three performance metrics to track the learning process of FeederGAN in terms of network connectivity and phase-transition continuity. The first metric checks the connectivity of generated networks. An adjacency matrix is *connected* when the matrix contains fully connected graph without isolated partitions. So an adjacency matrix with isolated blocks will fail the connectivity check because it causes the generated graph to break into a few disconnected pieces, in which case we consider the generated feeder infeasible. Once a graph is connected, we further exam the phase transition. The second and third metrics, *Success* and *Perfect* as defined in Table II below, check for phase-transition continuity that determines how the *3-phase, 2-phase and 1-phase* circuits shall connect to each other. For example, only "*a*", "*b*" or "*ab*" line segments can be branched out from an "*ab*" line segment.

Recall that when applying the feeder-head permutation method, we have, in total, $m$ possible topologies for each reconstructed adjacency matrix $\tilde{A}$. Thus, we calculate the performance *rate* using the number of graphs meeting the metric (i.e., *connected*, *success* or *perfect*) divided by $m$.

Table II Definition of the Success and Perfect metrics

| | Success | Perfect | |
|---|---|---|---|
| phase | subsequent neighbor's phase | phase | subsequent neighbor's phase |
| 3-phases | 3-phases, 2-phases, 1-phase | *abc* | *abc, ab, ac, bc, a, b, c* |
| | | *ab* | *ab, a, b* |
| | | *ac* | *ac, a, c* |
| 2-phases | 2-phases, 1-phase | *bc* | *bc, b, c* |
| | | *a* | *a* |
| 1-phase | 1-phase | *b* | *b* |
| | | *c* | *c* |

## D. Baseline Comparison

To quantify the benefit of the adversarial learning of using the FeederGAN-based process, we constructed a baseline case called the random-matrix based approach for comparison. In the random-matrix based approach, the Generator will randomly generate $\hat{A}$ and $\hat{X}$ and no Discriminator is used for comparing the generated topologies with the actual topologies. The same reconstruction rules are applied to make the depth of the resulting topology much larger than the breadth.

To compare the random-matrix based approach with the FeederGAN approach, we average the results obtained form 500 trials and present the performance metrics and the feasibility rate in Table III. We can see from the results that the adversarial structure improves the connected rate by 18.9%, the success and perfect rate by 27.2%,

Table III Performance comparison with and without adversarial learning

| | Random Matrix | FeederGAN |
|---|---|---|
| Connected rate | 8.3% | 27.2% |
| Success rate | 0 | 27.2% |
| Perfect rate | 0 | 27.2% |

## E. Mode Collapse and Human Guidance

Furthermore, we record the performance during the training process. After every 50 Generator iterations, we output a set of generated graphs and calculate the performance rates. Thus, in a 500-iteration window, we can obtain 10 rates and we average the 10 values to obtain the average feasibility rate for the 500-iteration window. This is because the rates may vary up and down because only a snapshot of the iteration is used to calculate them.

Thus, for 5,000 iterations, 10 averaged feasibility rates are obtained. As shown in the left plot in Fig. 12, the model performance improves to about 25% after 4,000 iterations, representing that about 25% of the $m$ generated feeder topologies are connected, success, and perfect.

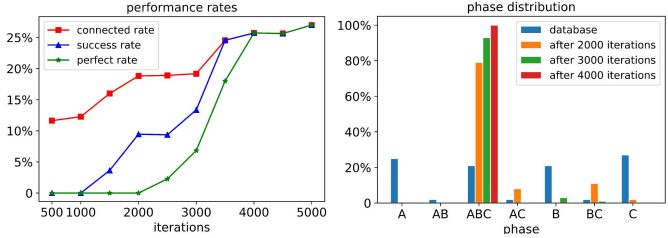

Fig. 12. Performance rates (left) and phase mode collapse (right)

During the training, we observed two major problems related with mode collapse [29], where the Generator locked into one or a few limited modes to exploit the Discriminator. The first is phase mode collapse. As shown in the right plot in Fig. 12, at the beginning of the training, 3-phase and single phase circuits are mixed evenly. However, after 2,000 iterations, 3-phase circuits start to dominate with the presence of a small number of 2-phase (i.e., $ac$ and $bc$) circuits; while towards the end, all circuits converge to 3-phase.

To solve this problem, the early-stop-strategy is used to terminate the training after approximately 3,000 to 4,000 iterations, when the phases generated are not all 3-phase circuits. Then, we use the human guidance to assign the phase information to line segments as illustrated in Algorithm 6. The advantage of applying human guidance are twofold: achieve a user-defined phase diversity and convert previously imperfect generated circuit topology to a perfect one.

**Algorithm 6. Human Guidance for Phase Assignment.**
1: Traverse the feeder from the feeder head with 3-phase attribute, i.e., $abc$;
2: Keep the phase attribute until reaching a bifurcating point;
3: Check for possible phase options via perfect definition in Table II;
4: Select from the possible choices the one with the highest score in the generated *softmax* approximated phase results, i.e. $\hat{X}_{cat}$.

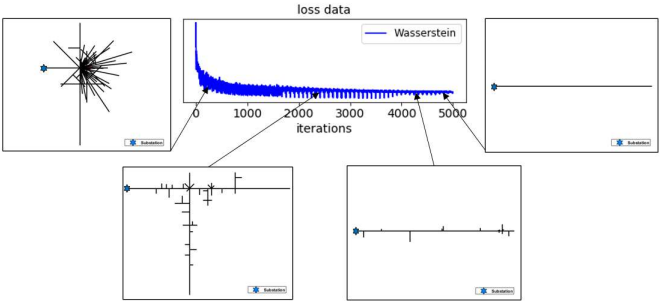

Fig. 13. Topology mode collapse

The second is topology mode collapse. We have plotted a few graphs to show the topologies of a few generated circuits at different locations in the Wasserstein loss curve, as shown in Fig. 13. At the very beginning, the feeder generated is dense and has shorter backbones. After 2000 iterations, the feeder shows a longer backbone with branches. But after 4000 iterations, the backbone remains but the branches disappear, showing a single line without bifurcation. This problem can be addressed by an early stop. In our experiments, we stop the training between 3,000 and 4,000 iterations to achieve satisfactory results.

### F. Quality Screening and Empirical Statistics

Empirical statistics can be used to measure realisticness. From real distribution feeder models (See Table IV), statistical characteristics describing the connectivity and phase relations of realistic feeders can be obtained. By comparing with those empirical statistics, we can discard generated feeders with characteristics outside the empirical ranges.

The first realisticness measurement is "Level" that measure the feeder topology hierarchy from the feeder head to the lowest downward load node. If a generated feeder has more than 10 levels, it is discarded. The second realisticness measure is the ratio of 3-phase, 2-phase, and 1-phase circuits. This is because a realistic feeder usually has more 1-phase circuits and 3-phase circuits with only a few being 2-phase circuits. The third realisticness measure is the out-degree of a node that defines the number of its subsequent neighbors. For a realistic feeder topology, 99% nodes has an out-degree less than 5. Moreover, as shown in Fig. 14, if the probability distribution of the length of generated line segments differs significantly from a real one, we can either discard it or use the probability distribution functions of real feeders to conduct a $2^{nd}$ round "length" re-generation to make sure that the length of each generated line segment is realistic.

Table IV Empirical statistics

| Metrics | Empirical Statistics | | | | | | |
|---|---|---|---|---|---|---|---|
| Level | $4\sim 7$ | | | | | | |
| Phase distribution | $a$ | $ab$ | $abc$ | $ac$ | $b$ | $bc$ | $c$ |
| | 18%~28% | 1%~3% | 20%~25% | 1%~3% | 18%~28% | 1%~3% | 18%~28% |
| Out-degree distribution | 0 | 1 | 2 | 3 | 4 | $\geq 5$ | |
| | 20%~40% | 25%~45% | 18%~26% | 5%~7% | 1%~3% | <1% | |

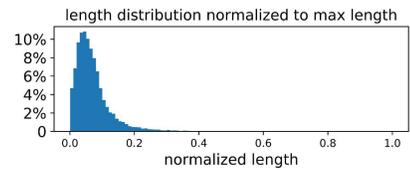

Fig. 14. Probability distribution function of the line segments

### VII. CONCLUSION AND FUTURE WORK

In this paper, we introduced FeederGAN, a novel GAN-based method, to enable automated, high fidelity synthetic feeder generation. FeederGAN ingests power system distribution feeder models as directed graphs using a device-as-node representation. Feeder topology and device characteristics are coded into the adjacent matrix and the attribute matrix to allow GCN-based methods to learn topology and attributes from actual power system feeder model input files. This approach will make FeederGAN *expandable* (handling feeders at different scales with any number of customized attributes) and having superb learning flexibility (learning from a full network structure or substructures). To solve the mode collapse problem, *early stop* and *human guidance* are proposed. The realisticness of the generated feeder topology can be validated by visual inspection and empirical statistics. Our simulation results have shown that even an expert cannot distinguish the generated topologies from the real ones. By comparing the empirical statistics, the generated synthetic feeders match well with the actual. This research aims at meeting the research needs of researchers for conducting distribution system planning studies and for developing control and energy management algorithms, where a large amount of realistic test feeders are needed. Our future work will focus on automated device placement so that sectionalizers and voltage regulation devices can be automatically added to the feeder model.

## Appendix

**Example 1**: This example illustrates the reconstruction process via feeder head permutation. As shown in Fig. A1, the reconstructed $\tilde{A}$ has only 1 non-zero entry in each column with only row 1, 2 and 3 has non-zero entries. Hence, nodes 1, 2 and 3 are the candidates for the feeder head node. If node 1 is chosen, set all entries in column 1 as 0. Traverse all the nodes to build the directed graph, i.e. $1 \rightarrow 2, 2 \rightarrow 3/4/5$. Then, convert a node-to-edge transformation so a feeder topology is obtained following the power system conventions.

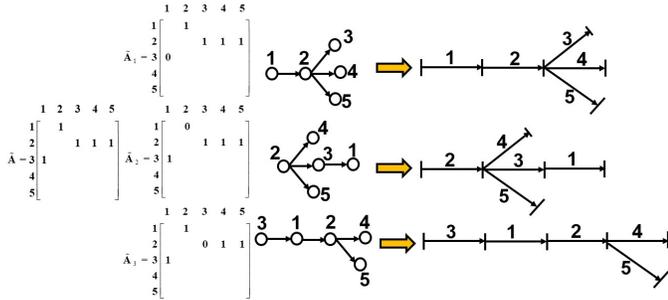

Fig. A1. Topology Reconstruction

**Example 2**: In Fig.A2, we illustrate how to calculate the pseudo coordinates. Suppose we have already traversed to bus $j$ with a coordinate at $(x_j, y_j)$ and a branch direction of $\theta$. The bifurcation leads to 4 children devices. The device linking bus $j$ and $k$ with fewest downward devices will have the lowest priority ranking with an angle assigned as $\delta$. Hence, for this branch, the direction angle is updated as $\theta + \delta$. The coordinate of bus $k$ is $(x_k, y_k)$, calculated by $x_k = x_j + l \cdot \cos(\theta + \delta)$, $y_k = y_j + l \cdot \sin(\theta + \delta)$, where $l$ is the device length between bus $j$ and $k$.

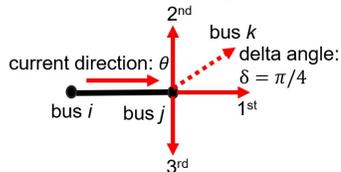

Fig. A2. Pseudo Coordinates calculation

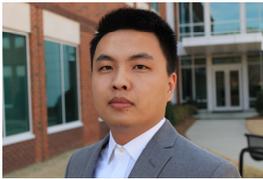

**Ming Liang** is a Ph.D. candidate at Future Renewable Electric Energy Delivery and Management (FREEDM) Systems Center, Dept. of Electrical and Computer Engineering, North Carolina State University (NCSU). He received M.S. in Electrical Engineering from NCSU in 2019 and B.E. in Electrical Engineering and Automation from Chongqing University in 2014. His research interests include renewable energy integration, load modeling and non-intrusive load monitoring. He is currently working on synthetic testing system and data generation in electric power distribution systems.

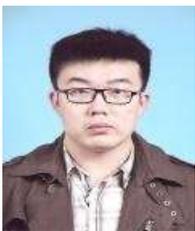

**Yao Meng** received the B.S. degree from Huazhong University of Science and Technology, Wuhan, China, and the M.S. degree from Zhejiang University, Hangzhou, China, in 2012 and 2015, respectively. In 2019, he received the Ph.D. degree from North Carolina State University, Raleigh, U.S.A.. He is now with China Electric Power Planning and Engineering Institute. His major research interests include distribution system analysis, optimal planning and operation of power systems with energy storage.

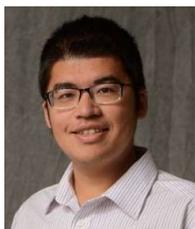

**Jiyu Wang** received the B.S. degree in electrical engineering from China Agricultural University, Beijing, China, in 2014, and the M.S. degree in electrical engineering from North Carolina State University, Raleigh, NC, USA, in 2016, where he is currently pursuing the Ph.D. degree. His research interests include distribution feeder modeling, solar integration studies, and demand response.

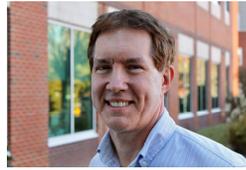

**David L. Lubkeman** (IEEE Fellow) received the Ph.D. degree in electrical engineering from Purdue University with an emphasis in power systems engineering. He is also a Licensed Professional Engineer. Dr. Lubkeman is currently a Research Professor at North Carolina State University. His research interests include electric power distribution system analysis, distribution automation, distribution management systems, renewable energy integration and microgrid applications.

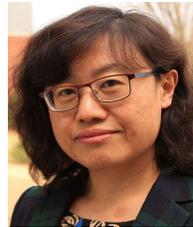

**Ning Lu** (M'98--SM'05) received the B.S. degree in electrical engineering from the Harbin Institute of Technology, Harbin, China, in 1993, and the M.S. and Ph.D. degrees in electric power engineering from Rensselaer Polytechnic Institute, Troy, NY, USA, in 1999 and 2002, respectively. Dr. Lu is currently a professor with the Electrical and Computer Engineering Department, North Carolina State University. She was with Pacific Northwest National Laboratory. Her current research interests include microgrid modeling and control, power distribution systems operation and planning, and power system data analysis.